\documentclass[twocolumn, twocolappendix]{aastex631}
\usepackage{amsmath,bm}
\usepackage{multirow}
\usepackage{enumitem}
\usepackage{color}
\usepackage{ulem}
\usepackage{CJKfntef,CJK}
\usepackage{makecell,ulem}

\newcommand\hcop{$\rm HCO^+$}
\newcommand\coa{$\rm ^{12}CO$}
\newcommand\cob{$\rm ^{13}CO$}
\newcommand\coc{$\rm C^{18}O$}
\newcommand\hcn{$\rm HCN$}
\newcommand\kms{$\rm km\ s^{-1}$}
\newcommand\IhcopIhcn{$I({\rm HCO^+})/I({\rm HCN})$}
\newcommand\NhcopNco{$N({\rm HCO^+})/N({\rm CO})$}

\begin{document}
\begin{CJK*}{UTF8}{gbsn}
\title{Mapping the dense molecular gas towards thirteen supernova remnants}

\author[0000-0002-9776-5610]{Tian-yu Tu (涂天宇)}
\affiliation{School of Astronomy \& Space Science, Nanjing University, 163 Xianlin Avenue, Nanjing 210023, China}

\author[0000-0002-4753-2798]{Yang Chen (陈阳)}
\affiliation{School of Astronomy \& Space Science, Nanjing University, 163 Xianlin Avenue, Nanjing 210023, China}
\affiliation{Key Laboratory of Modern Astronomy and Astrophysics, Nanjing University, Ministry of Education, Nanjing 210023, China}
\email{ygchen@nju.edu.cn}

\author[0000-0002-5786-7268]{Qian-Cheng Liu (刘前程)}
\affiliation{School of Astronomy \& Space Science, Nanjing University, 163 Xianlin Avenue, Nanjing 210023, China}

\begin{abstract}
Supernova remnants (SNRs) can exert strong influence on molecular clouds (MCs) through interaction by shock wave and cosmic rays. 
In this paper, we present our mapping observation of \hcop\ and \hcn\ 1--0 lines towards 13 SNRs interacting with MCs, together with archival data of CO isotopes. 
Strong \hcop\ emission is found in the fields of view (FOVs) of SNRs W30, G9.7$-$0.0, Kes\,69, 3C\,391, 3C\,396, W51C, HC\,40, and CTB109 in the local-standard-of-rest (LSR) velocity intervals in which they are suggested to show evidence of SNR-MC interaction. 
We find an incomplete \coa\ shell surrounding G9.7$-$0.0 with an expanding motion. 
This shell may be driven by the stellar wind of the SNR progenitor. 
We also find an arc of \coa\ gas spatially coincident with the northwestern radio shell of Kes\,69. 
As for the \hcop\ line emission, SNRs 3C\,391 and W51C exhibit significant line profile broadening indicative of shock perturbation, and CTB109 exhibits a possible blue-shifted line wing brought by shock interaction. 
We do not find significant variation of the \IhcopIhcn\ line ratio between broad-line and narrow-line regions, among different SNRs, and between MCs associated with SNRs and typical Galactic MCs. 
Therefore, we caution on using the \IhcopIhcn\ line ratio as a diagnostic of SNR feedback and CR ionization. 
We also estimate the \NhcopNco\ abundance ratio in 11 regions towards the observed SNRs, but they show little difference from the typical values in quiescent MCs, possibly because \NhcopNco\ is not an effective tracer of CR ionization. 
\end{abstract}

\keywords{Molecular clouds (1072) --- Supernova remnants (1667)}

\section{Introduction} \label{sec:intro}
Massive ($\gtrsim 8M_{\odot}$) stars, born in molecular clouds (MCs) and ending their lives as supernovae (SNe), each re-deposit $\sim 10^{51}\rm \ erg$ back to the interstellar medium and regulate the galactic ecology and evolution \citep{Pillepich_Simulating_2018}. 
Many supernova remnants (SNRs) are found to be located in the vicinity of MCs \citep[e.g.][]{Jiang_Cavity_2010,Chen_Molecular_2014}. 
If an SNR is interacting with an MC, the shock, X-ray emission, and the cosmic rays (CRs) accelerated by the SNR can strongly affect the physical and chemical properties of the MC \citep{Vink_Physics_2020}, which can be observed in terms of molecular transitions. 

\par

CO and its isotopes are the most popular molecular species employed to investigate the SNR-MC interaction \citep[e.g.,][]{Zhou_Systematic_2023} which can be evidenced by morphology alignment of the CO emission with the SNR radio emission, and the line broadening or line wings \citep[e.g.,][]{Zhou_Discovery_2009a,Su_Discovery_2009,Jiang_Cavity_2010,Zhou_INTERACTION_2016}. 
However, the CO molecules can only trace the most extended and low-density (with critical density\footnote{The definition of critical density of a molecular transition is not consistent in a variety of the literature. Here we define it as the Einstein $A$ coefficient divided by the collisional rate coefficient at a specific temperature. The coefficients are taken from the Leiden Atomic and Molecular Database (LAMDA): \url{https://home.strw.leidenuniv.nl/~moldata/}.} of $\sim 2\times 10^3\rm \ cm^{-3}$ for the 1--0 line and $3\times 10^4\rm \ cm^{-3}$ for the 3--2 line at 10 K) part of the MCs. 
Although denser shocked gas with broadened line profiles has been found in several SNRs \citep[e.g.,][]{Koo_Interaction_1997,Reach_Excitation_1999,Reach_Shocked_2005,Lee_Identification_2012}, mapping observations of dense molecular gas around SNRs still lack. 

\par

The \hcop\ and \hcn\ 1--0 lines, with critical densities of $\sim 2\times 10^5\rm \ cm^{-3}$ and $10^6 \rm \ cm^{-3}$, respectively, are typical dense gas tracers in MCs \citep{Shirley_Critical_2015}. 
The integrated intensity ratio of the two lines (hereafter called line ratio for simplicity) have been used as diagnostics of physical processes in extragalactic studies \citep[e.g.][]{Krips_Multi-Transition_2008,Costagliola_Molecules_2011}. 
A simultaneous mapping of CO and these two dense gas tracers can provide a comprehensive view of the molecular environment of SNRs. 

SNRs are believed to be the main accelerators of Galactic CRs \citep[e.g.,][]{Aharonian_Gamma_2013}. 
The high-energy (with kinetic energy $\gtrsim280$ MeV) CR protons can generate $\gamma$-ray emission originated from the decay of $\pi^0$ mesons produced in the collision between the high-energy CR protons and H nuclei in MCs, which is called the hadronic scenario. 
The high-energy CR electrons can produce $\gamma$-ray emission via the inverse Compton scattering of background radiation, which is called the leptonic scenario. 
The $\gamma$-ray emission can be detected by instruments such as the \textit{Fermi} Large Area Telescope \citep{Acero_FIRST_2016}. 
On the other hand, the low-energy CR protons serve as the main source of ionization in dark MCs shielded from UV radiation \citep{Padovani_Cosmic-ray_2009}. 
This process starts the formation of poly-atomic molecules in MCs and regulates the molecular chemistry in MCs. 

\par

Direct estimation of CR ionization rates in MCs associated with SNRs typically uses the \NhcopNco\ and $N({\rm DCO^+})/N({\rm HCO^+})$ abundance ratios \citep{Ceccarelli_Supernova-enhanced_2011,Vaupre_Cosmic_2014}. 
Although the emission of $\rm DCO^+$ is hard to detect, both observations obtained enhanced CR ionization rates.
The enhanced CR ionization rates induced by SNRs can also be revealed indirectly by various observations, including 1720 MHz OH masers and 6.4 keV Fe I K$\alpha$ line. 
The production of the 1720 MHz OH masers requires enhanced CR ionization rate in addition to SNR-MC interaction \citep{Lockett_OH_1999,Wardle_Supernova_2002}. 
The Fe I K$\alpha$ line is produced via inner-shell ionization of neutral iron when protons in the MeV band collide with MCs \citep{Nobukawa_Evidence_2018}. 
Detection of hadronic $\gamma$-ray emission also hints that low-energy CR protons are possibly also accelerated and ionizing the MCs. 

\par

The \hcop\ molecule has been expected to be a tracer of CR ionization rate in MCs \citep{Ceccarelli_Supernova-enhanced_2011,Bayet_Chemistry_2011,Albertsson_Atlas_2018}. 
Specifically, enhanced abundance of \hcop\ relative to CO due to CR ionization in shocked MCs has been found in SNRs W49B and W28 \citep{Zhou_Unusually_2022b,Tu_Shock_2024}
Therefore, observation of \hcop\ in MCs associated SNRs may provide information about the CR acceleration and ionization in SNRs.\par
In this paper, we performed new mapping observations in \hcop\ and \hcn\ 1--0 lines, supplemented by archival data of CO isotopes, towards a sample of 13 SNRs suggested to be interacting with MCs in order to obtain the spatial distribution of dense molecular gas around them, search for shocked dense gas with line profiles broadened, and study the viability of \IhcopIhcn\ line ratio as a diagnostic of SN feedback and CR ionization. 

\par

The paper is structured as follows. 
In Section \ref{sec:obs}, we describe our new observations and other archival data. 
We present the observational results in Section \ref{sec:res} and discuss the SNR-MC interaction, the \IhcopIhcn\ line ratio, and the \NhcopNco\ abundance ratio in Section \ref{sec:disc}. 
Finally, a summary is given in Section \ref{sec:con}. 

\section{Observations} \label{sec:obs}
\subsection{\hcop\ and \hcn\ observations and data reduction}

Mapping observations of the 1--0 emission lines of \hcop\ and \hcn\ were performed with the 13.7m millimeter-wavelength telescope of the Purple Mountain Observatory at Delingha (hereafter PMOD; PI: Y.\ Chen). 
We included thirteen SNRs with evidence of SNR-MC interaction in our observations: W30, G9.7$-$0.0, G16.7$+$0.1, Kes\,69, Kes\,75, 3C\,391, Kes\,78, 3C\,396, 3C\,397, W51C, and CTB109 \citep[according to][]{Jiang_Cavity_2010}. 
The \hcop\ and \hcn\ lines were simultaneously observed in 2016--2022. 
The Fast Fourier Transform Spectrometers with 1 GHz bandwidth and 16,384 channels were used as the back ends, providing a velocity resolution of 0.21 \kms\ at 89 GHz. 
The coverage of the observations is shown in Table \ref{tab:info} where we also list some basic information about the target SNRs.
The half-power beamwidth (HPBW) of the telescope at 89 GHz is $\approx 60^{\prime\prime}$. 
The main beam efficiencies (within 0.58--0.63 across the years of abservation) were corrected according to the annual status reports of PMOD\footnote{\url{http://www.radioast.nsdc.cn/zhuangtaibaogao.php}}. 
The raw data were reduced with the GILDAS/CLASS package\footnote{\url{https://www.iram.fr/IRAMFR/GILDAS/}}. 
The data cubes of \hcop\ and \hcn\ were all resampled to have the same velocity channel width of 0.25 \kms\ and the same pixel size of $30^{\prime\prime}$. 
The RMS noise measured in main beam temperature ($T_{\rm mb}$) is $\sim 0.05$ K for CTB109 and $\sim0.1$ K for the other SNRs.

\subsection{Other archival data}
Additional data of the 1--0 line of CO isotopes were retrieved from FOREST Unbiased Galactic plane Imaging survey with the Nobeyama 45-m telescope \citep[FUGIN,][]{Umemoto_FOREST_2017} and Milky Way Image Scroll Painting \citep[MWISP,][]{Su_Milky_2019}. 
The FUGIN project covers $|b|\leq 1^\circ$, $10^\circ\leq l\leq50^\circ$ and $198^\circ\leq l\leq236^\circ$. 
It has an angular resolution of $\approx 20^{\prime\prime}$, a sensitivity of $\sim 1\text{--}3 \rm \ K$ at a velocity channel width of 0.65 \kms\ for \coa, and a sensitivity of $\sim 0.6\text{--}1.5 \rm \ K$ at a velocity channel width of 0.65 \kms\ for \cob\ and \coc. 
The MWISP project has covered $|b|\leq 5^\circ$ and $-10^\circ\leq l\leq+250^\circ$ with the 1--0 lines of \coa, \cob, and \coc\ and has an angular resolution of $\approx 50^{\prime\prime}$, a sensitivity of $\sim 0.5$ K at a velocity channel of 0.16 \kms\ for \coa, and a sensitivity of $\sim 0.3$ K at a velocity channel of 0.17 \kms\ for \cob\ and \coc. 
Specifically, we used the MWISP data for SNRs W30, G9.7$-$0.0, HC\,40, and CTB109, while the FIGIN data for the other SNRs. 
All the data cubes of CO lines were smoothed to an angular resolution of $60^{\prime\prime}$ so as to have better comparison with the \hcop\ and \hcn\ data. 

\par

Data of the radio continuum of the SNRs were taken from the ``SNRcat" SNR catalog\footnote{\url{http://snrcat.physics.umanitoba.ca}} \citep{Ferrand_census_2012} and the VLA Galactic Plane Survey \citep[VGPS,][]{Stil_VLA_2006} project to delineate the boundary of the SNRs. 

\par

All the processed data were further analyzed with \emph{Python} packages Astropy \citep{AstropyCollaboration_Astropy_2018,AstropyCollaboration_Astropy_2022} and Spectral-cube \citep{Ginsburg_Radio_2015}.
The data cubes were reprojected with Montage\footnote{\url{http://montage.ipac.caltech.edu/}} package when necessary. 
We visualized the data with \emph{Python} package Matplotlib\footnote{\url{https://matplotlib.org/}}.

\begin{deluxetable*}{ccccc}
\tablecaption{Basic information of target SNRs.\label{tab:info}}

\tablehead{ \colhead{Name} & \colhead{Map size} & \colhead{Velocity of interaction (Reference)} & \colhead{\makecell{CR acceleration$^{\rm a}$ \\ (Reference)}} & \colhead{\makecell{\hcop \\ detection$^{\rm b}$}} }
\startdata
\makecell{G8.7$-$0.1 \\ (W30)} & $29^\prime\times 28^\prime$ & \begin{tabular}[c]{@{}c@{}}OH maser: $+36$ \kms\ (1)\\ molecular lines: $+8$--$+56$ \kms\ (2)\end{tabular} & OH (1), H$\gamma$ (24) & Y \\  \hline
G9.7$-$0.0 & $24^\prime\times 20^\prime$ & OH maser: $+43$ \kms\ (1) & OH (1), PH$\gamma$ (25) & Y \\ \hline
G16.7$+$0.1 & $19^\prime\times 19^\prime$ & \begin{tabular}[c]{@{}c@{}}OH maser: $+20$ \kms\ (3)\\ CO 1--0: $+25.1$--$+25.9$ \kms\ (4)\\ broad CO 2--1: $+25$ \kms\ (5)\end{tabular} & OH (3), U$\gamma$ (26) & N \\ \hline
\makecell{G21.8$-$0.6 \\ (Kes\,69)} & $28^\prime\times 28^\prime$ & \begin{tabular}[c]{@{}c@{}}compact OH maser: $+69$ \kms\ (3) \\ extended OH maser: $+85$ \kms\ (6) \\ CO 1--0: $\sim +85$ \kms\ (7) \end{tabular} & OH (3, 6), FeK (27) & Y \\ \hline
\makecell{G29.7$-$0.3 \\ (Kes\,75)} & $19^\prime\times 19^\prime$ & \begin{tabular}[c]{@{}c@{}}CO 1--0: $+45$--$+58$ \kms\ (8) \\ broad CO 2--1: $+53$ \kms\ (5) \end{tabular} & L$\gamma$ (28) & N \\ \hline
\makecell{G31.9$+$0.0 \\ (3C\,391)} & $18^\prime\times 18^\prime$ & \begin{tabular}[c]{@{}c@{}}OH maser: $+105$ and $+110$ \kms\ (9) \\ broad molecular lines: $+105$ \kms\ (10) \\ high-$J$ CO lines: $\sim +105$ \kms\ (11) \end{tabular} & \makecell{OH (9), FeK (29), \\ H$\gamma$ (30), Chem (31)} & Y \\ \hline
\makecell{G32.8$-$0.1 \\ (Kes\,78)} & $29^\prime\times 29^\prime$ & \begin{tabular}[c]{@{}c@{}}OH maser: $+86$ \kms\ (12) \\ CO 1--0 and 2--1: $\sim +81$ \kms\ (13) \end{tabular} & \makecell{OH (12), FeK (27), \\ U$\gamma$ (32)} & P \\ \hline
\makecell{G39.2$-$0.3 \\ (3C\,396)} & $21^\prime\times 21^\prime$ & \begin{tabular}[c]{@{}c@{}}CO 1--0 and 2--1: $\sim +84$ \kms\ (14) \\ broad CO 2--1: $\sim +69$ and $\sim +77$ \kms\ (5) \end{tabular} & H$\gamma$ (33) & Y \\ \hline
\makecell{G41.1$-$0.3 \\ (3C\,397)} & $19^\prime\times 19^\prime$ & \begin{tabular}[c]{@{}c@{}}CO 1--0: $\sim +32$ \kms\ (15) \\ broad CO 2--1: $\sim +31$ \kms\ (5) \end{tabular} & L$\gamma$ (34) & N \\ \hline
\makecell{G49.2$-$0.7 \\ (W51C)} & $19^\prime\times 19^\prime$ & \begin{tabular}[c]{@{}c@{}}OH maser: $+71.9$ and $+68.9$ \kms\ (3) \\ broad CO and $\rm HCO^+$: $+80$--$+120$ \kms\ (16) \\ CR ionization: $\sim +70$ \kms\ (17) \\ broad molecular lines: $\sim +70$ \kms\ (18) \\ SiO emission: $\sim +67$ \kms\ (19) \end{tabular} & \makecell{OH (3), FeK (35), \\ H$\gamma$ (36), Chem (17)} & Y \\ \hline 
\makecell{G54.4$-$0.3 \\ (HC\,40)} & $25^\prime\times 25^\prime$ & CO 1--0: $+36$--$+44$ \kms\ (20) & --- & Y \\ \hline
\makecell{G74.9$+$1.2 \\ (CTB87)} & $20^\prime\times 21^\prime$ & \begin{tabular}[c]{@{}c@{}}CO 1--0 and 3--2: $\sim -58$ \kms\ (21) \\ CO 1--0: $\sim -58$ \kms\ (22) \end{tabular} & L$\gamma$ (37) & N \\ \hline
\makecell{G109.1$-$1.0 \\ (CTB109)} & \begin{tabular}[c]{@{}c@{}}N$^{\rm c}$: $34^\prime\times 18^\prime$\\ W$^{\rm b}$: $17^\prime\times 25^\prime$\end{tabular} & CO 1--0: $\sim -55$ \kms\ (23) & PH$\gamma$ (38) & Y \\ 
\enddata
\tablecomments{
$^{\rm a}$ Signature of enhanced CR ionization rate. OH: Detection of 1720 MHz OH maser. FeK: Detection of 6.4 keV Fe I K$\alpha$ line. Chem: molecular chemistry induced by enhanced CR ionization rate. H$\gamma$: Confirmation of hadronic $\gamma$-ray emission. L$\gamma$: Confirmation of leptonic $\gamma$-ray emission. PH$\gamma$: Detection of $\gamma$-ray which is possibly due to the hadronic scenario. U$\gamma$: Detection of $\gamma$-ray whose origin has not been discussed. \\
$^{\rm b}$ Whether \hcop\ emission is detected. Y: Strong emission of \hcop\ is detected. P: Weak \hcop\ emission is detected with limited spatial extent. N: No \hcop\ is detected in the FOV of the SNR. \\
$^{\rm c}$ Two regions of CTB109 are mapped and coadded. \\
References: 
1. \citet{Hewitt_Discovery_2009}, 
2. \citet{Feijen_Arcminute-scale_2020}, 
3. \citet{Green_Continuation_1997}, 
4. \citet{Reynoso_CO_2000}, 
5. \citet{Kilpatrick_Systematic_2016}, 
6. \citet{Hewitt_Survey_2008b}, 
7. \citet{Zhou_Discovery_2009a}, 
8. \citet{Su_Discovery_2009}, 
9. \citet{Frail_Survey_1996}, 
10. \citet{Reach_Excitation_1999}, 
11. \citet{Gusdorf_molecular_2014}, 
12. \citet{Koralesky_Shock-excited_1998}, 
13. \citet{Zhou_Molecular_2011}, 
14. \citet{Su_Molecular_2011}, 
15. \citet{Jiang_Cavity_2010}, 
16. \citet{Koo_Interaction_1997}, 
17. \citet{Ceccarelli_Supernova-enhanced_2011}, 
18. \citet{Brogan_Discovery_2006}, 
19. \citet{Dumas_Localized_2014}, 
20. \citet{Junkes_54_1992a}, 
21. \citet{Kothes_Distance_2003}, 
22. \citet{Liu_Investigation_2018a}, 
23. \citet{Sasaki_Evidence_2006}, 
24. \citet{Liu_GeV_2019}, 
25. \citet{Yeung_Studying_2016}, 
26. \citet{Acero_FIRST_2016},
27. \citet{Nobukawa_Evidence_2018}, 
28. \citet{Straal_Discovery_2023}
29. \citet{Sato_Discovery_2014}, 
30. \citet{Ergin_Recombining_2014}, 
31. \citet{Tu_Yebes_2024}, 
32. \citet{Auchettl_Fermi-LAT_2014}, 
33. \citet{Sezer_Suzaku_2020}, 
34. \citet{Bhattacharjee_Investigating_2017}, 
35. \citet{Shimaguchi_Suzaku_2022a}, 
36. \citet{Jogler_Revealing_2016}, 
37. \citet{Saha_Implications_2016}, 
38. \citet{Castro_Fermi-LAT_2012}.
}
\end{deluxetable*}

\section{Results} \label{sec:res}
\subsection{Observational results of \hcop\ and \hcn} \label{sec:res_hcop}

\begin{figure*}
\centering
\includegraphics[height=0.85\textheight]{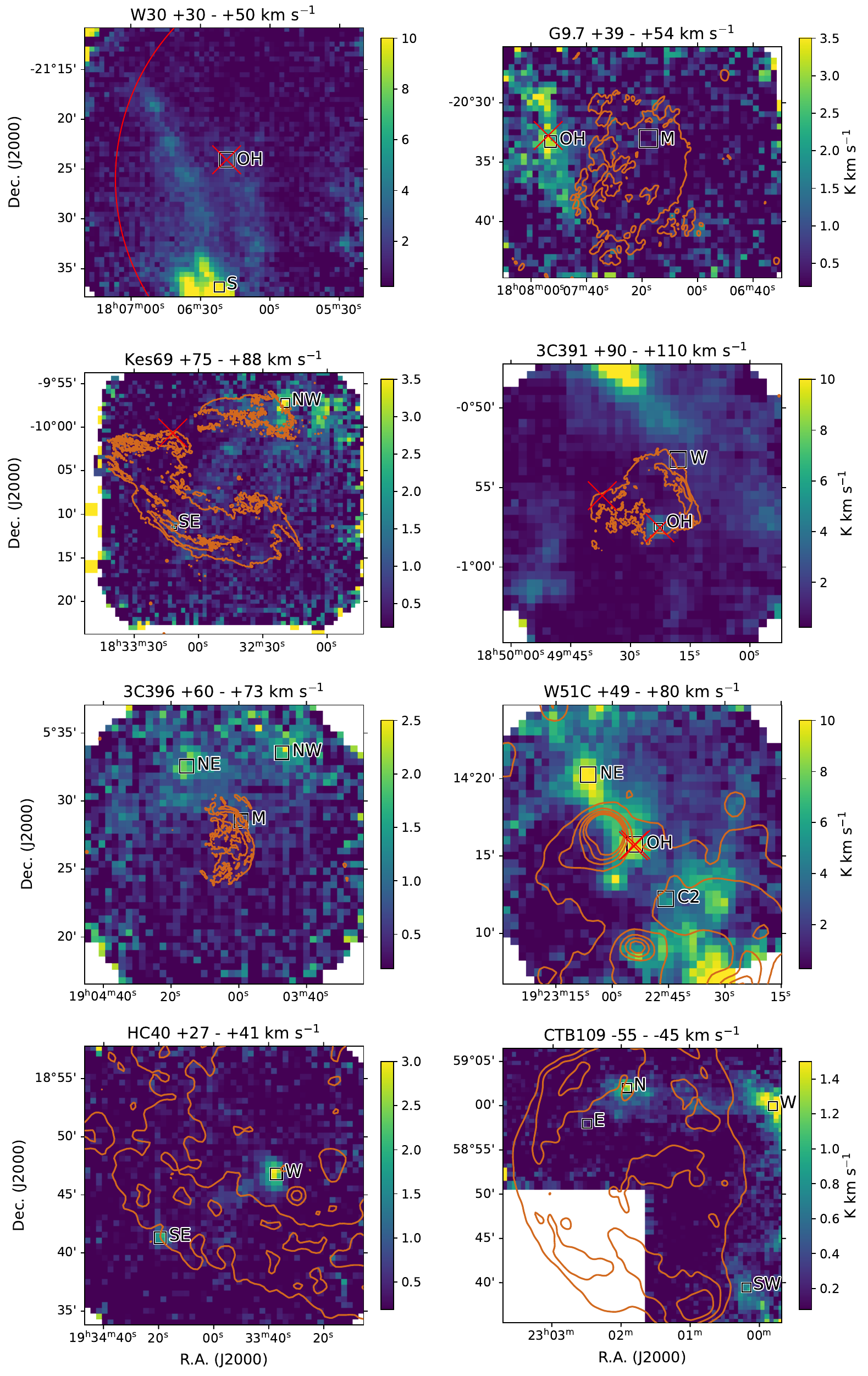}
\caption{Integrated intensity maps of \hcop\ in the 8 SNRs with strong \hcop\ emission. The spectrum extraction portions and the LSR velocity ranges of the integration are labeled on top of each of the panels. The orange contours show the radio continuum. For W30, we delineate the boundary of the SNR with a red arc. The red crosses shows the 1720 MHz OH masers. The black boxes mark the regions where we extract molecular spectra. 
\label{fig:HCOp_map}}
\end{figure*}

\begin{figure*}
\centering
\includegraphics[height=0.9\textheight]{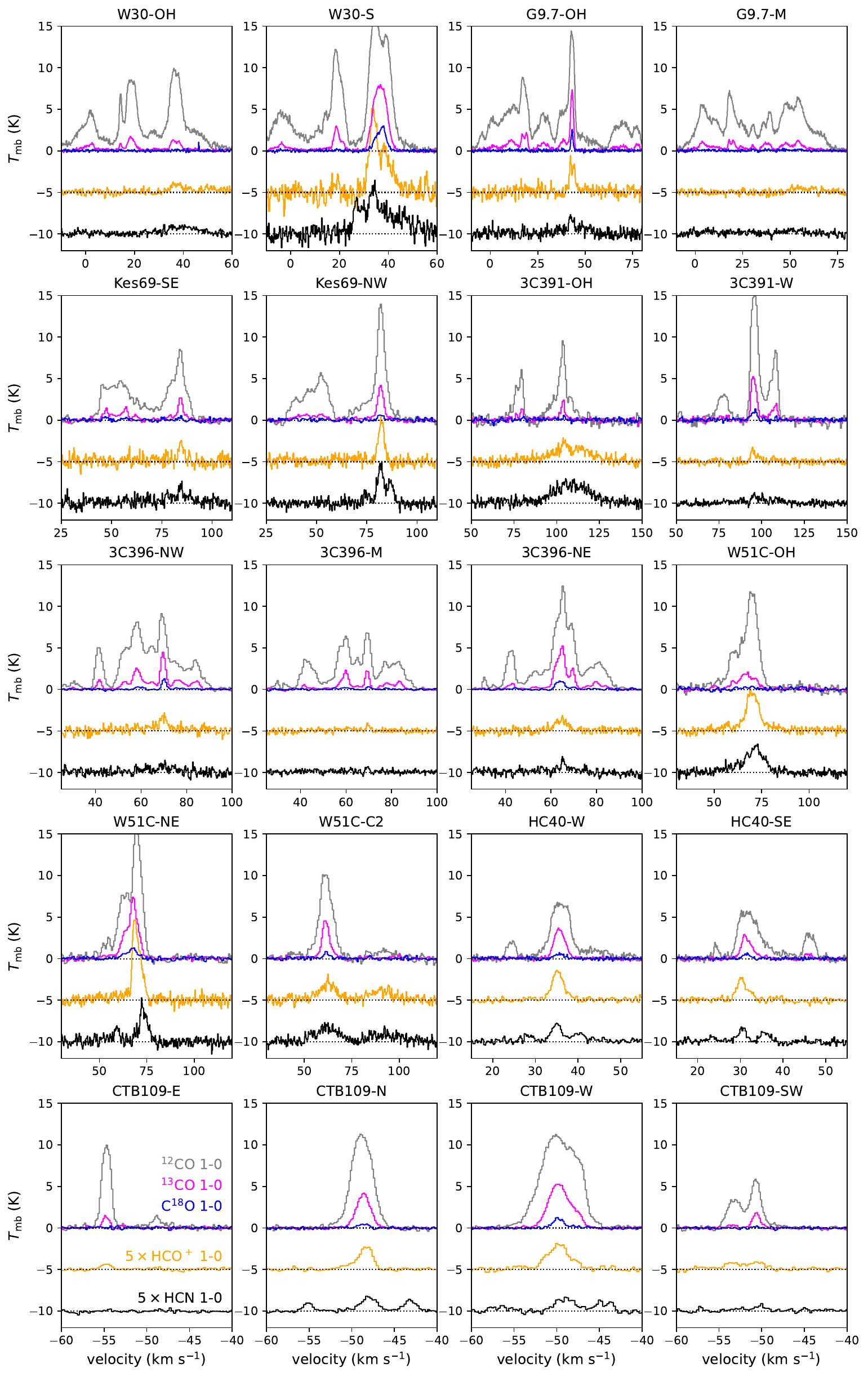}
\caption{\coa\ 1--0 (grey), \cob\ 1--0 (magenta), \coc\ 1--0 (blue), \hcop\ 1--0 (orange), and \hcn\ 1--0 (black) spectra averaged in the black boxes in Figure \ref{fig:HCOp_map}. The \hcop\ and \hcn\ spectra are multiplied by a factor of 5 and lowered by 5\,K and 10\,K, respectively, for better visualization. 
\label{fig:spec}}
\end{figure*}

The local-rest-of-rest (LSR) velocity intervals in which the observed SNRs are interacting with MCs are listed in Table \ref{tab:info}. 
We detected strong emission of \hcop\ in the fields of view (FOVs) of SNRs W30, G9.7$-$0.0, Kes\,69, 3C\,391, 3C\,396, W51C, HC\,40, and CTB109 in the velocity intervals in which previous studies have found evidence of SNR-MC interaction. 
In Kes\,78, we only detected weak \hcop\ emission, while no \hcop\ and \hcn\ emission was found in SNRs G16.7$+$0.1, 3C\,397, Kes\,75, and CTB87 in corresponding velocity intervals. 

\par

In Figure \ref{fig:HCOp_map}, we display the integrated intensity maps of \hcop\ in the FOVs of the 8 SNRs with strong \hcop\ emission. 
The spatial distribution of \hcn\ is similar in large to that of \hcop. 
The LSR velocity ranges are consistent with those suggested to exhibit SNR-MC interaction by previous studies. 
The spectra extracted from the regions marked by black boxes in Figure \ref{fig:HCOp_map} are shown in Figure \ref{fig:spec}. 
Here we briefly describe the observational results of the morphology and spectra of the \hcop\ and \hcn\ emission in the 8 SNRs and compare our results with previous studies. 

\par

\textit{G8.7$-$0.1 (W30)}: 
Our map of W30 is centered on the 1720 MHz OH maser discovered by \citet{Hewitt_Discovery_2009}, which is located in the eastern part of the SNR. 
The boundary of W30 shown in Figure \ref{fig:HCOp_map} was taken from the Green's catalogue of Galactic SNRs\footnote{\url{https://www.mrao.cam.ac.uk/surveys/snrs/snrs.info.html}}.
The emission of \hcop\ mainly consists of three parts: a stripe extending southward from the OH maser, another brighter stripe located to the east of the first stripe, and a very bright source containing several HII regions \citep{Feijen_Arcminute-scale_2020} in the southern edge of the map. 
Towards the OH maser, the spectrum of \hcop\ consists of two components centered at $\approx +36$ and $\approx +38$ \kms, consistent with the \cob\ 1--0 line, while in the bright source W30-S, the two components of \hcop\ are located at $\approx +34$ and $\approx +40$ \kms. 
All of the velocities are roughly consistent with the systemic velocity of the OH maser ($+36$ \kms), suggesting that the dense gas traced by \hcop\ is located at the same distance as W30 or even is the same cloud that harbors the maser. 

\par

\textit{G9.7$-$0.0} (hereafter G9.7 for short): 
Extended \hcop\ emission is located to the east of G9.7 and covers the position of the OH maser. 
The line profile of \hcop\ towards the OH maser exhibits double peak feature, the dip velocity of which is similar to the peak position of \cob\ and \coc\ lines at $\approx +43.4$ \kms. 
But at other positions, the systemic velocity of \hcop\ is the same as that of \cob\ and \coc\ ($\sim +43$ \kms, not shown). 
We report a marginal detection of \hcop\ inside the SNR (G9.7-M). 
The seemingly broad line profile of \hcop\ and \hcn\ may be due to different components as shown in the \cob\ line. 

\par

\textit{G21.8$-$0.6 (Kes\,69)}: 
\hcop\ emission of Kes\,69 is mainly located on the northwestern boundary of the SNR at $\approx +83$ \kms. 
Small-scale \hcop\ emission is also found towards the southeastern shell of Kes\,69 (see region Kes\,69-SE). 
The systemic velocity of the \hcop\ and \hcn\ lines ($\approx +85$ \kms) towards Kes\,69-SE is consistent with both of that of the extended OH maser found by \citet{Hewitt_Survey_2008b} and that of the CO 1--0 and \hcop\ obtained by \citet{Zhou_Discovery_2009a}. 
We do not detect \hcop\ and \hcn\ emission associated with the OH maser towards the northeast of the SNR at $\sim +69$ \kms\ which is the systemic velocity of the 1720 MHz OH maser found by \citet{Green_Continuation_1997}. 

\par

\textit{G31.9$+$0.0 (3C\,391)}: 
SNR 3C\,391 is believed to be interacting with a large MC located towards its west side, which creates the bright radio bar along its western boundary \citep[e.g.][]{Reach_Molecular_2002}. 
We find weak \hcop\ emission in this direction (e.g., 3C\,391-W), which extends northeastward to reach a star-forming region \citep{Urquhart_ATLASGAL_2018} with a strong enhancement in the northern edge of the map. 
Broadened lines of \hcop\ and \hcn\ are found towards the southern OH maser (i.e. 3C\,391-OH), which is consistent with the results of \citet{Reach_Excitation_1999}.
We do not detect \hcop\ towards the northeastern OH maser. 

\par

\textit{G39.2$-$0.3 (3C\,396)}: 
Bright \hcop\ emission in the field of view is located outside the northern boundary of the SNR, e.g. region 3C\,396-NE at $\approx +66$ \kms\ and region 3C\,396-NW at $\approx +70$ \kms. 
Their large distances from the SNR suggest that they have not yet been hit by the SNR shock.
Weak emission of \hcop\ and \hcn\ is detected inside the radio boundary (e.g., region 3C\,396-M) at $\approx +70$ \kms, which is consistent with the results of \citet{Kilpatrick_Systematic_2016} who found broadened \coa\ 2--1 line to the north of the SNR at $+69$ \kms. 
However, multi-wavelength study suggested that 3C\,396 is associated with a MC at $\sim +84$ \kms\ \citep{Su_Molecular_2011}. 

\par

\textit{G49.2$-$0.7 (W51C)}: 
Our map of SNR W51C is centered at the 1720 MHz OH masers \citep{Green_Continuation_1997}, where (see region W51C-OH) there is a bright \hcop\ emission spot. 
However, \citet{Tian_High-velocity_2013} argued that the OH masers may result from the HII region, G49.2$-$0.3, which exhibits bright radio continuum and is located on the east of the OH masers. Another \hcop\ spot (W51C-NE) is located to the north of the HII region. 
Previous studies have found broad molecular lines at both $\sim +70$ \kms\ \citep{Brogan_OH_2013} and $\sim +90$ \kms\ \citep{Koo_Interaction_1997}. 
Our results show that the \hcop\ emission is aligned from northeast to southwest across the field of view. 
Most of the emission is centered at $\sim +70$ \kms, but the line profiles do not exhibit broadened feature at this LSR velocity. 
The systemic velocities of \hcop\ and \hcn\ towards W51C-NE show a slight difference, and we will further discuss this in Section \ref{sec:hcophcn}. 
We also find broadened \hcop\ and \hcn\ lines at $\sim +90$ \kms\ towards the positions of clump~2 and clump~4 (the former is abbreviated as W51-C2 in Figure \ref{fig:HCOp_map}) reported in \citet{Koo_Interaction_1997}. 

\par

\textit{G54.4$-$0.3 (HC\,40)}: 
Our observation covers the southeastern part of SNR~HC\,40. 
The noticeable emission of \hcop\ are concentrated in two clumps: HC\,40-SE and HC\,40-W. 
HC\,40-SE is consistent with a protostellar clump G54.373$-$0.614 at +31 \kms\ \citep{Urquhart_ATLASGAL_2018}. 
The asymmetric broad line profile of \hcop\ may be the result of two crowded velocity components. 
HC\,40-W is located at a systemic velocity of $\approx +35$ \kms\ and does not exhibit any evidence of star-formation \citep{Urquhart_ATLASGAL_2018}. 
Weak \hcop\ emission extends from HC\,40-W towards southeast. 

\par

\textit{G109.1$-$1.0 (CTB\,109)}: 
The \hcop\ emission in the mapped field of SNR CTB\,109 is mainly located in the north, northwest, and southwest of the SNR (see regions CTB109-N, CTB109-NW, and CTB109-SW) and at the LSR velocities centered at $-55$ -- $-45$ \kms. 
Weak \hcop\ emission connects regions CTB109-N and CTB109-W. 
\citet{Sasaki_Evidence_2006} found shocked \coa\ 1--0 line towards the ``lobe" (approximately CTB109-E in our figure), but we find only weak \hcop\ emission and no evidence of line broadening. 
However, we find a blue-shifted line wing towards CTB109-N, which may be due to the shock interaction. 
Similar line profile can also be seen, though not that remarkable, in the \coa\ emission. 
Towards CTB109-W, the \hcop\ line, similar to the \coa\ and \cob\ lines, consists of two velocity components. 

\subsection{CO towards G9.7 and Kes\,69}\label{sec:res_co}
\begin{figure*}[htbp]
\centering
\includegraphics[width=0.99\textwidth]{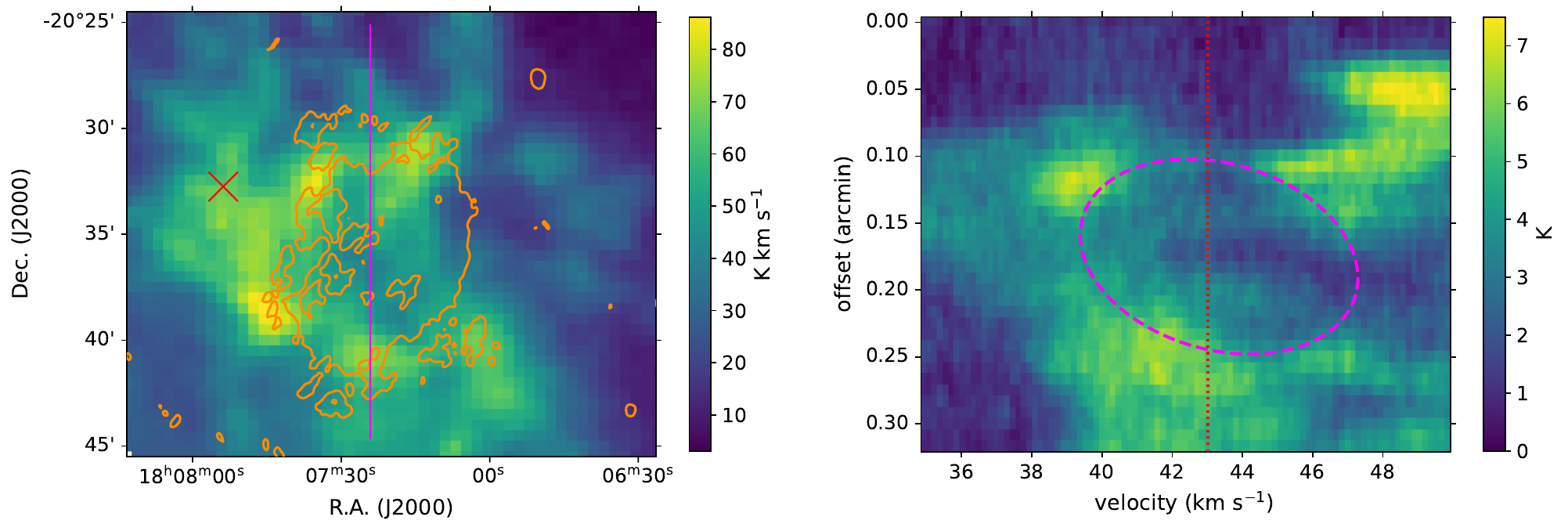}
\caption{\textit{left panel}: Integrated intensity map of \coa\ towards G9.7 between +39 and +54 \kms, overlaid with 327 MHz radio continuum (the level is 40 mJy/beam). The red cross shows the OH maser detected by \citet{Hewitt_Discovery_2009}, and the magenta line shows the path along which we extract the position-velocity map. 
\textit{right panel}: Position-velocity map of \coa\ along the magenta line in the left panel (start from north). The dashed magenta ellipse delineates the possible bubble structure, while the vertical dotted red line shows the systemic velocity of the OH maser (+43 \kms). 
\label{fig:g9.7}}
\end{figure*}

\begin{figure}[htbp]
\centering
\includegraphics[width=0.47\textwidth]{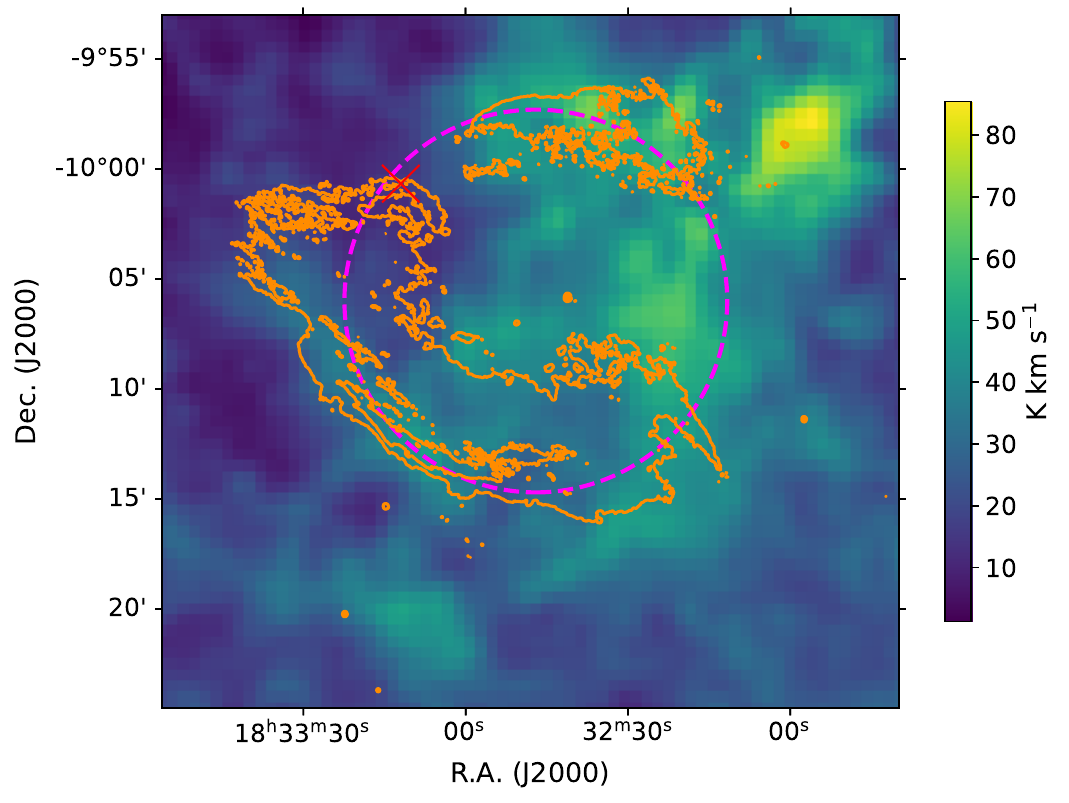}
\caption{Integrated intensity map of \coa\ towards Kes\,69 between +75 and +85 \kms, overlaid with 1420 MHz radio continuum (the levels are 2 and 8 mJy/beam). The red cross shows the OH maser detected by \citet{Green_Continuation_1997}. The dashed magenta circle is similar to the circle in Figure 5 of \citet{Zhou_Discovery_2009a} which shows roughly the molecular arcs associated with Kes\,69.
\label{fig:Kes69}}
\end{figure}

In this section, we briefly introduce our new results on the spatial distribution of CO gas towards SNRs G9.7 and Kes\,69. 

\par

\textit{G9.7}: In the left panel of Figure \ref{fig:g9.7}, we show the integrated intensity map of \coa\ towards G9.7. 
Compared with \hcop\ (See \S\ref{sec:res_hcop} and Figure \ref{fig:HCOp_map}), the emission of \coa\ is much more extended. 
Besides the \coa\ gas spatially coincident with the \hcop\ emission, we also find an ``C''-shaped incomplete \coa\ shell coincident with the radio continuum of G9.7 in the northern, eastern and southern parts. 
The right panel of Figure \ref{fig:g9.7} shows the position-velocity (PV) diagram along the magenta line shown in the left panel. 
As marked by the dashed magenta ellipse, an annular pattern is located within a velocity between +39 and +47 \kms\ and an offset within $0^\prime.10$--$0^\prime.25$, which is often related to the expanding motion of gas \citep[e.g.][]{Zhou_Expanding_2016}. 

\par

\textit{Kes\,69}: Figure \ref{fig:Kes69} shows the integrated intensity map of \coa\ towards Kes\,69. 
Although \citet{Zhou_Discovery_2009a} have discussed the \coa\ emission associated with Kes\,69, their data of radio continuum suffers from lower sensitivity and does not reveal the radio arc towards the northwestern boundary but only shows a subtle hint there. 
With the new radio data, we find that an short arc of \coa\ emission, together with more compact \hcop\ emission (Kes\,69-NW in Figure \ref{fig:HCOp_map}), is spatially coincident with the northwestern radio arc of Kes\,69. 

\section{Discussion} \label{sec:disc}

\subsection{SNR-MC interaction}
Evidence of SNR-MC interaction includes morphological agreement of molecular lines with SNR emission in radio or X-ray band, molecular line broadening or asymmetric line profile, 1720 MHz OH masers, line emission with high high-to-low excitation line ratio, etc. \citep{Jiang_Cavity_2010}. 
In this section, we investigate the first two observational facts in our sample of SNRs. 

\subsubsection{Morphology agreement of \coa\ molecular features with SNRs G9.7 and Kes\,69}
G9.7 was identified by \citet{Brogan_Discovery_2006} as an SNR, and was confirmed to be interacting with MCs by the detection of a 1720 MHz OH maser at +43 \kms\ \citep{Hewitt_Discovery_2009}. 
However, no detailed molecular line observation was conducted before. 
In the left panel of Figure \ref{fig:g9.7}, we show an incomplete \coa\ shell surrounding the radio boundary of G9.7 (\S\ref{sec:res_co}). 
This suggests that the \coa\ shell may be associated with the SNR. 
The \coa\ bubble, expanding at a velocity of $\sim 4$ \kms, is likely to be driven by the stellar wind of the SNR progenitor. 
This scenario can be seen in a number of molecular bubbles surrounding SNRs, such as Tycho's SNR \citep{Zhou_Expanding_2016}, VRO 42.05.01 \citep{Arias_environment_2019}, and Kes\,67 (Shen et al., submitted), with expanding velocities around 5 \kms.  

\par

We note that the OH maser detected by \citet{Hewitt_Discovery_2009} seems to be located outside the radio boundary. 
This may be due to the limited sensitivity of the radio image, since the data of \citet{Brogan_Discovery_2006} shows that the 327 MHz radio continuum extends towards the northeastern side and approaches the position of the OH maser. 
The radio morphology in \citet{Brogan_Discovery_2006} also exhibits an agreement with the \hcop\ emission, suggesting possible interaction. 

\par

SNR Kes\,69 was found to be interacting with the MCs by the detection of both compact OH maser \citep{Green_Continuation_1997} and extended OH maser \citep{Hewitt_Survey_2008b}. 
The CO emission was reported by \citet{Zhou_Discovery_2009a} who identified a \coa\ arc coincident with the radio arc located in the southeastern boundary of Kes\,69. 
They also plotted a \coa\ arc located towards the northwest of Kes\,69, but limited by the sensitivity of the radio data, the coincidence of the arc and the radio shell was confused with an HII region G21.902$-$0.368. 
The new radio data clearly shows a radio shell towards the northwest of Kes\,69, making up a circle (similar to that in Figures 4 and 5 in \citet{Zhou_Discovery_2009a}) together with the southeastern radio shell (see Figure \ref{fig:Kes69}). 
Considering that the northwestern \coa\ emission is located at the similar velocity of the southeastern \coa\ shell ($\sim +81$ \kms), 
we suggest that, in addition to the known molecular shell along the southeastern boundary
of the SNR, there is also a molecular arc along the northwestern boundary.

\subsubsection{Line broadening and asymmetric line profile} \label{sec:disc_broadening}
Molecular line broadening and asymmetric line profile have also been regarded as strong evidence of SNR-MC interacting \citep{Jiang_Cavity_2010}. 
This criterion has been widely used in the diagnostics with CO lines: asymmetric \coa\ line together with non-detection of \cob\ has been used to identify the SNR-cloud interaction in SNRs W44 \citep{Seta_Detection_2004}, 3C\,396 \citep{Su_Molecular_2011}, HB3 \citep{Zhou_INTERACTION_2016}, G51.26$+$0.11 \citep{Zhong_study_2023a}, etc. 
However, the CO lines may suffer from line crowding, especially in the inner Galaxy, where multiple emission components of the \coa\ 1--0 line crowd in a small velocity interval, which would confuse the line broadening due to SNR-MC interaction. 

\par

Compared with the \coa\ 1--0 line, the \hcop\ 1--0 line traces a denser part of molecular cloud and suffer less from line crowding. 
The \hcop\ 1--0 transition is easy to get optically thick \citep{vanderTak_computer_2007}, which allows it to be broadened by shock disturbance. 
Therefore, broadened \hcop\ 1--0 can also serve as a shock tracer. 
Among the 13 SNRs we observed, we find shocked \hcop\ towards 3C\,391 and W51C (3C\,391-OH and W51C-C2, see Figure \ref{fig:spec}), consistent with the results of \citet{Reach_Excitation_1999} and \citet{Koo_Interaction_1997}. 

\par

In addition to these two SNRs known to exhibit broadened \hcop\ line, a blue-shifted line wing of \hcop\ towards the CTB109-N region is also noticeable. 
This region is compatible with the X-ray absorption reported by \citet{Sasaki_XMM-Newton_2004}. 
If the MC is absorbing the X-ray emission of the SNR, it could be located in front of the SNR. 
The SNR shock interacting with the MC could drives the molecular gas to move towards us, which results in a blue-shifted wing in molecular transitions, as we can see in the \hcop\ 1--0 line. 
However, we could still not rule out the possibility that the line wing is caused by another component due to the limited signal-to-noise ratio (S/N) of our data. 
Further high-sensitivity observation is needed to confirm the interaction between CTB109 and its adjacent dense molecular gas. 

\par

In the FOVs of other SNRs, we do not detect any shock broadening or line wings brought by the SNRs. 
However, our observations are limited in sensitivity, and the $\sim 1^\prime$ beam size may also lead to severe beam dilution, both of which could make it difficult for us to detect the shocked \hcop\ and \hcn\ emission. 
According to \citet{Zhou_Unusually_2022b}, the peak main beam temperature $T_{\rm peak}$ of the shocked \hcop\ emission can be as low as $\sim 0.04$ K. 
Therefore, our observations could have missed some shocked dense gas in the SNRs. 
Besides, for young SNRs, it is possible that the shock-cloud interaction has too short timescale to have caused detectable line broadening \citep{Fukui_Pursuing_2021}. 

\par

The 1720 MHz OH maser is strong signpost of SNR-MC interaction \citep[e.g.,][]{Jiang_Cavity_2010}. 
This OH satellite line ($^2\Pi_{3/2},\ J=3/2,\ F=2\rightarrow1$) maser is pumped by collisions with other molecular species \citep{Wardle_Supernova_2002}. 
The pumping of these masers requires a moderately high density ($\sim 10^5\rm \ cm^{-3}$), warm temperature ($\sim$50--125 K), and high column density of OH ($\sim 10^{-16}$--$\rm 10^{-17} \rm \ cm^{-2}$) \citep{Lockett_OH_1999}. 
The required density is in favor of the excitation of \hcop\ and \hcn. 
However, in our observation, although significant \hcop\ emission is found towards regions W30-OH, G9.7-OH, 3C\,391-OH, and W51C-OH, no \hcop\ emission is spatially coincident with the OH masers towards G16.7$+$0.1, the northeast of Kes\,69, the northeast of 3C\,391, and Kes\,78. 
Although this may be due to the limited sensitivity of our observation, a more naturally explanation is that the OH masers are rather compact \citep[$\sim10^2\text{--}10^3\rm \ AU$ compared with the $1^\prime$ beam of our observation which corresponds to $\sim 1$ pc at a distance of 4 kpc,][]{Hoffman_OH_2005} and only reflect the physical conditions in a very small region. 
Therefore, it is unreasonable to use the pumping condition of the OH masers to represent the physical properties of the entire MC.

\begin{figure*}
\centering
\includegraphics[width=0.90\textwidth]{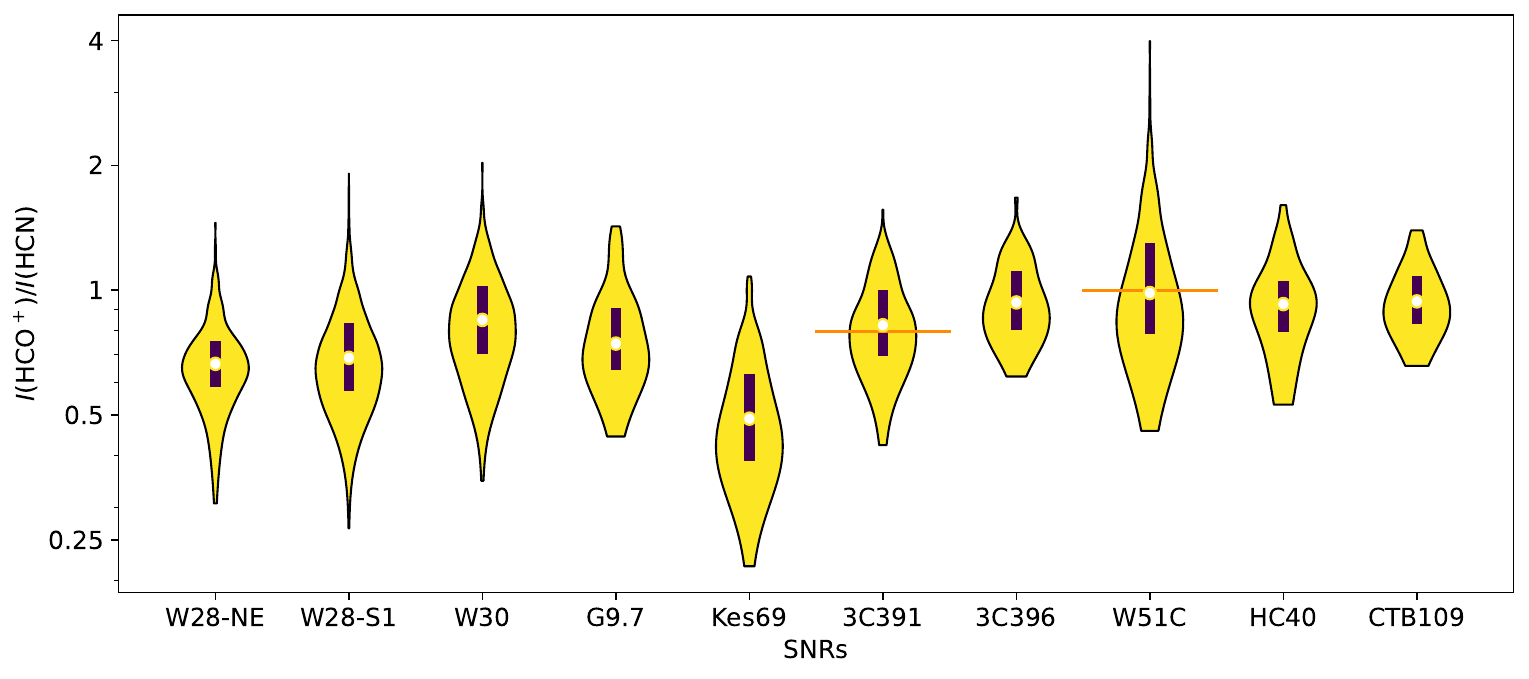}
\caption{Violin plot showing the ranges of line ratio \IhcopIhcn\ in the entire FOVs of the 8 SNRs and two regions of SNR W28. Mean values are shown with white dots while quartiles are shown as the two ends of each thick black bar. 
The two horizontal orange lines show the \IhcopIhcn\ of the broadened lines towards 3C\,391-OH and W51C-C2. 
\label{fig:violin}}
\end{figure*}

\subsection{\IhcopIhcn\ line ratio} \label{sec:hcophcn}

In Figure \ref{fig:violin}, we display the violin plot showing the ranges of line ratio \IhcopIhcn\ in the FOVs of the 8 SNRs with significant \hcop\ emission. 
The velocity ranges for the integration are chosen so as to cover all the \hcop\ and \hcn\ molecules with suggested interaction with the SNRs, and only points with a $\rm S/N>5\sigma$ are included. 
We also show the line ratios towards W28-NE and W28-S1 (See \citet{Tu_Shock_2024} and Fugure 1 therein for the definition of the two small regions: W28-NE is the shocked MC towards SNR W28, while W28-S1 is a complex of MCs exhibiting star-forming activities free from the disturbance of W28). 
The values of \IhcopIhcn\ towards 3C\,391-OH and W51C-C2, i.e. the two shocked MCs with broadened \hcop\ and \hcn\ lines, are shown in horizontal orange lines. 

\par
As can be seen from the figure, the median values of \IhcopIhcn\ in all of the SNRs fall in 0.65--1.0, except for Kes\,69 of which the median value is 0.5. 
The scatters of the \IhcopIhcn\ ratios (estimated as the difference between the quartiles and the median values) are all within 0.2, except for W51C of which the upper quartile is 0.3 greater than the median value. 
The \IhcopIhcn\ ratios exhibit little difference in shocked (broadened line, in W28-NE, 3C\,391-OH and W51C-C2) and unshocked (narrow line) clouds, and dose not show significant variation among SNRs. 

\begin{figure}
\centering
\includegraphics[width=0.45\textwidth]{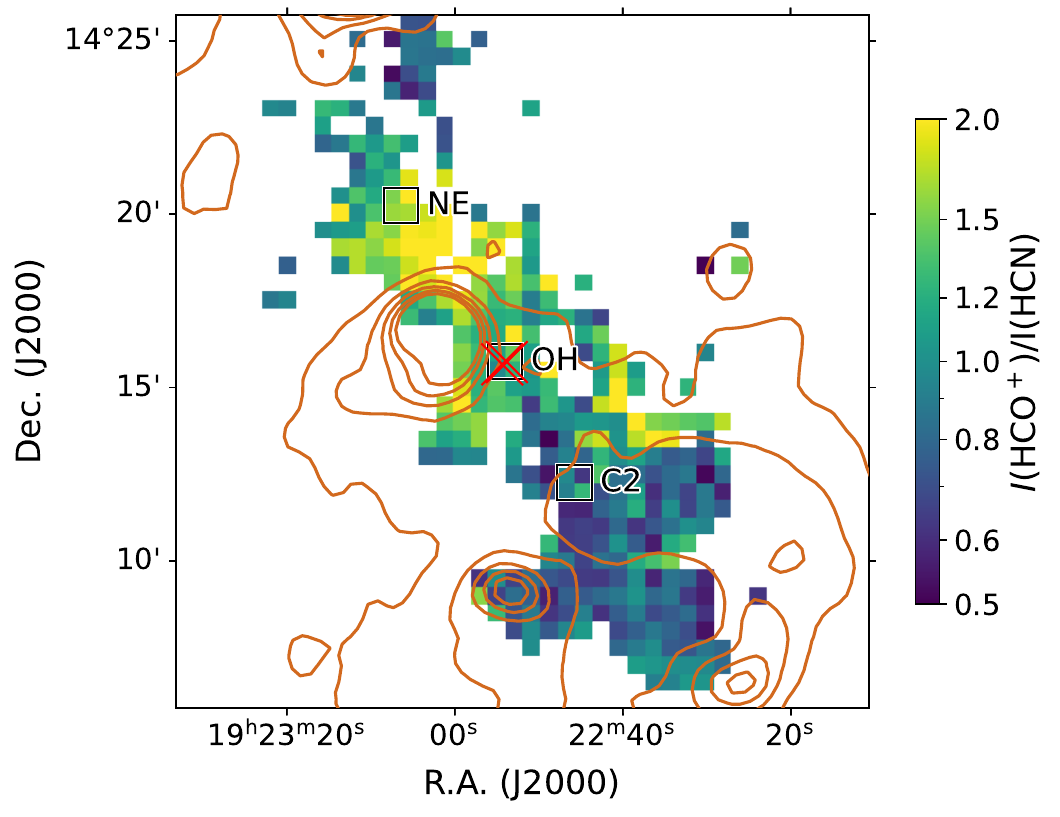}
\caption{\IhcopIhcn\ line ratio map towards SNR W51C, overlaid with contours of VGPS radio continuum in steps of 50, 100, 150, 200, 250 K. The red crosses show the 1720 MHz OH masers detected by \citet{Green_Continuation_1997}, and the black boxes are the regions where we extract the spectra plotted in Figure \ref{fig:spec}. 
\label{fig:W51Cratio}}
\end{figure}

\begin{figure*}
\centering
\includegraphics[width=0.9\textwidth]{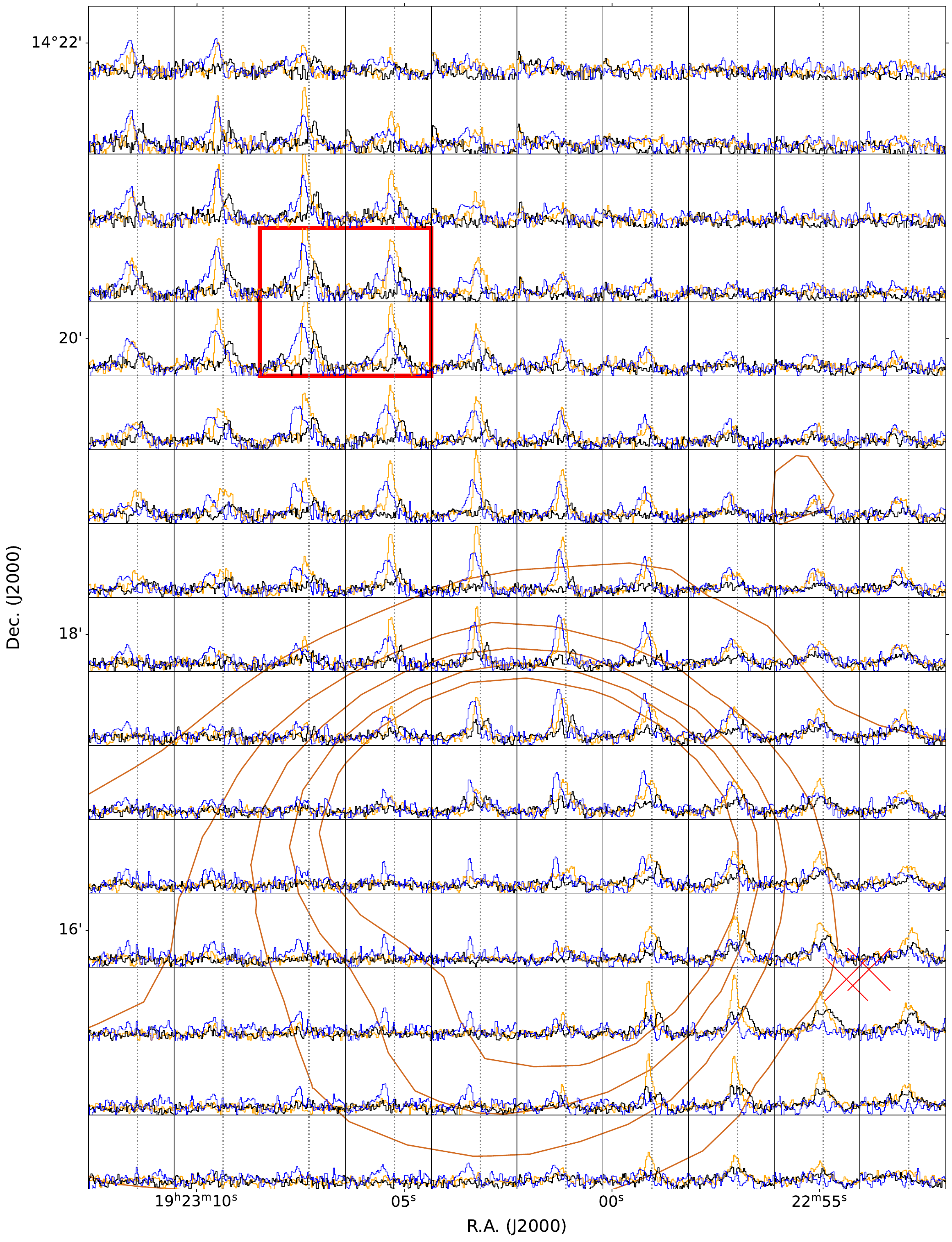}
\caption{Grid of \hcop\ (orange), \hcn\ (black) and \coc\ (blue) 1--0 line around W51C-NE restricted to a velocity range of +50 -- +85 \kms\ and a $T_{\rm mb}$ range of 0.2--2 K. The orange contours and red crosses are the same as Figure \ref{fig:W51Cratio}. The red box shows the W51C-NE region. The dotted vertical lines show the velocity of +70 \kms\ for reference. 
\label{fig:W51Cgrid}}
\end{figure*}

The observed region of SNR W51C exhibits the largest dispersion and the highest values of \IhcopIhcn. 
In Figure \ref{fig:W51Cratio}, we show the \IhcopIhcn\ line ratio map towards W51C. 
It is clear that the line ratio is higher towards the northeastern and lower towards the southwestern. 
To investigate why \IhcopIhcn\ is enhanced towards the northeastern part of the observed region, we take a closer look at the spectra of W51C-NE, and make a grid of \hcop, \hcn\ and \coc\ spectra in Figure \ref{fig:W51Cgrid}.

\par

From the spectra of W51C-NE (Figure \ref{fig:spec}), we find that the spectrum of \hcop\ is red-shifted compared with \coc, while \hcn\ is further red-shifted compared with \hcop. 
Both \hcop\ and \hcn\ exhibit asymmetric line profile, with a sharp decrease towards the blue side. 
Figure \ref{fig:W51Cgrid} shows that the velocity shift of \hcop\ and \hcn\ pervades the entire region with enhanced \IhcopIhcn\ line ratio. 
For the \hcn\ line, we notice a weak component at $\approx+59$ \kms\ in W51C-NE, which is not consistent with any other components shown in other molecular transitions. 
The high \IhcopIhcn\ and velocity shift could be explained by self-absorption of \hcop\ and \hcn, which is stronger for the \hcn\ line. 
The self-absorption is blue-shifted so that the resulting spectra are red-shifted. 
In this case, the +59 \kms\ component corresponds to an unabsorbed hyperfine structure of \hcn. 
This scenario is similar to the blue-profile of contracting molecular cloud cores \citep{Evans_Physical_1999a}, but the ``red-profile'' in our case indicates expansion \citep[e.g.][]{Mardones_Search_1997}. 
The expanding motion could be due to oscillation of the MC triggered by external perturbation \citep{Fu_Starless_2011} like the heating of the HII region G49.2$-$0.3 with bright compact radio emission towards the south of W51C-NE \citep{Kim_Dense_2016}. 
We note that the stronger absorption of \hcn\ than \hcop\ is anomalous because the \hcop\ line tends to be more susceptible to self-absorption \citep{Aalto_Probing_2015}. 
A possible explanation of our result is that the absorbing layer has enhanced abundance of \hcn, leading to a stronger absorption. 

\par

The \IhcopIhcn\ line ratio has been studied in various astrophysical conditions. 
This value is $\sim 1.1$ in the Orion B giant molecular cloud (GMC) \citep{Santa-Maria_HCN_2023a}, $\sim 0.9$ in giant molecular filament 54 (GMF54) \citep{Wang_Dense_2020}, $\sim 1.32$ in infrared dark clouds (IRDCs) \citep[e.g.][]{Liu_systemic_2013}, $\sim 2$ in MCs in the outer Milky Way \citep{Braine_Dense_2023a}, $\sim 0.6$ in the central molecular zone (CMZ) of our Galaxy \citep[e.g.,][]{Jones_Spectral_2012,Santa-Maria_Submillimeter_2021}, $\sim 0.8$ in the star-forming disk of nearby massive spiral galaxies \citep{Jimenez-Donaire_EMPIRE_2019}, and varying from $\sim 0.4$ to $\sim 1.4$ in nearby active galaxies \citep[e.g.][]{Aladro_Lambda_2015}. 
Previous studies have found that mechanical heating by SNRs can lead to a high \IhcopIhcn\ line ratio at low density and low \IhcopIhcn\ at high density \citep{Loenen_Mechanical_2008,Costagliola_Molecules_2011,Privon_Dense_2017}.
High CR ionization rate, which is another factor brought by SNRs \citep[e.g.,][]{Ceccarelli_Supernova-enhanced_2011,Vaupre_Cosmic_2014}, can also enhance the \IhcopIhcn\ line ratio by altering the abundance ratio between \hcop\ and \hcn\ because the abundance of \hcop\ is expected to be enhanced with higher CR ionization rate \citep{Chin_Molecular_1997,Pound_CARMA_2018} while \hcn\ is not sensitive to CR ionization rate \citep{Albertsson_Atlas_2018,Holdship_History-independent_2022}. 
However, we do not find solid evidence of variation in \IhcopIhcn\ line ratio induced by SNRs from our observation. 
The difference in \IhcopIhcn\ between broad-line and narrow-line emission in W28, 3C\,391 and W51C is not significant, suggesting that the heating effect of the shock wave does not have a strong impact on \IhcopIhcn. 
Concerning CRs, which can diffuse or escape from the shock layer and induce ionization in narrow-line regions \citep{Ceccarelli_Supernova-enhanced_2011,Vaupre_Cosmic_2014}, we also fail to find prominent difference in \IhcopIhcn\ between SNRs and typical MCs, which does not support the enhancement of \IhcopIhcn\ by CRs. 

\par

We note, however, that CR ionization can affect the molecular abundances, but the relation between the abundance ratio and the line ratio is complicated. 
Besides the abundances of \hcop\ and \hcn, the line ratio is also affected by several physical factors such as gas density, temperature, infrared pumping, electron excitation, elemental abundance, etc. (see \citet{Salak_Dense_2018} and the reference therein).  
It is also possible that SNRs have similar effects on the brightness of \hcop\ and \hcn\ especially in shocked regions, since studies have suggested the chemical similarity between \hcop\ and \hcn\ \citep[e.g.][]{Liu_systemic_2013}. 
As a result, these effects may be eliminated when calculating the line ratio.

\par

\subsection{\NhcopNco\ abundance ratio} \label{sec:N}
\begin{deluxetable*}{cccccccc}
\tablecaption{Results of spectral decomposition and estimation of the abundance ratio \NhcopNco. \label{tab:N}}
\tablehead{
\colhead{Region} & 
\colhead{\makecell{$T_{\rm ex}$$^a$ \\ (K)}} & 
\colhead{Molecule} & 
\colhead{\makecell{$v_0$$^b$\\ (\kms)}} & 
\colhead{\makecell{$T_{\rm peak}$ \\ (K)}} & 
\colhead{\makecell{FWHM \\ (\kms)}} & 
\colhead{\makecell{$N$ \\ ($\rm cm^{-2}$)}} & 
\colhead{\NhcopNco$^c$} 
}
\startdata
\multirow{6}{*}{W30-S$^d$} & \multirow{3}{*}{18.9} 
   & \coa  & 34.29$\pm$0.05 & 15.44$\pm$0.18 & 5.12$\pm$0.09 & $2.5\times 10^{17}$ & \multirow{3}{*}{$3.2\times10^{-5}$}  \\
&  & \coc  & 34.95$\pm$0.22 &  1.43$\pm$0.13 & 3.09$\pm$0.29 & $6.4\times 10^{15}$ &  \\
&  & \hcop & 33.72$\pm$0.07 &  1.54$\pm$0.14 & 3.10$\pm$0.27 & $8.0\times 10^{12}$ &  \\
& \multirow{3}{*}{15.5} 
   & \coa  & 39.43$\pm$0.05 & 12.04$\pm$0.30 & 4.72$\pm$0.18 & $1.6\times 10^{17}$ & \multirow{3}{*}{$6.9\times10^{-5}$} \\
&  & \coc  & 37.88$\pm$0.12 &  2.73$\pm$0.12 & 3.21$\pm$0.16 & $1.2\times 10^{16}$ &  \\
&  & \hcop & 38.18$\pm$0.50 &  0.92$\pm$0.05 & 8.34$\pm$0.84 & $1.1\times 10^{13}$ &  \\ 
\hline
\multirow{3}{*}{G9.7-OH} & \multirow{3}{*}{16.6} 
   & \coa  & 43.13$\pm$0.02 & 13.20$\pm$0.17 & 2.82$\pm$0.04 & $1.1\times 10^{17}$ & \multirow{3}{*}{$3.3\times10^{-5}$} \\
&  & \coc  & 43.14$\pm$0.01 &  2.46$\pm$0.06 & 1.25$\pm$0.03 & $4.3\times 10^{15}$ &  \\
&  & \hcop & 43.28$\pm$0.04 &  0.68$\pm$0.04 & 3.53$\pm$0.22 & $3.6\times 10^{12}$ &  \\
\hline
\multirow{3}{*}{Kes\,69-SE} & \multirow{3}{*}{10.4} 
   & \coa  & 84.50$\pm$0.17 &  7.06$\pm$0.31 & 6.03$\pm$0.44 & $1.1\times 10^{17}$ & \multirow{3}{*}{$1.4\times10^{-5}$} \\
&  & \cob  & 84.64$\pm$0.10 &  2.57$\pm$0.17 & 3.14$\pm$0.23 & $9.9\times 10^{15}$ &  \\
&  & \hcop & 84.75$\pm$0.13 &  0.47$\pm$0.04 & 2.90$\pm$0.31 & $1.6\times 10^{12}$ &  \\
\hline
\multirow{3}{*}{Kes\,69-NW} & \multirow{3}{*}{15.1} 
   & \coa  & 82.44$\pm$0.04 & 11.64$\pm$0.28 & 4.14$\pm$0.12 & $1.3\times 10^{17}$ &\multirow{3}{*}{$3.8\times10^{-5}$} \\
&  & \cob  & 82.21$\pm$0.05 &  4.09$\pm$0.12 & 3.55$\pm$0.12 & $2.0\times 10^{16}$ &  \\
&  & \hcop & 82.32$\pm$0.06 &  0.99$\pm$0.03 & 3.44$\pm$0.14 & $4.9\times 10^{12}$ &  \\ 
\hline
\multirow{3}{*}{3C\,391-W} & \multirow{3}{*}{19.3} 
   & \coa  & 96.21$\pm$0.04 & 15.82$\pm$0.25 & 4.97$\pm$0.09 & $2.5\times 10^{17}$ & \multirow{3}{*}{$7.6\times10^{-6}$} \\
&  & \coc  & 95.90$\pm$0.18 &  1.09$\pm$0.10 & 4.07$\pm$0.42 & $6.4\times 10^{15}$ &  \\
&  & \hcop & 95.60$\pm$0.18 &  0.24$\pm$0.02 & 4.81$\pm$0.44 & $1.9\times 10^{12}$ &  \\ 
\hline
\multirow{3}{*}{3C\,396-NW} & \multirow{3}{*}{11.1} 
   & \coa  & 69.61$\pm$0.15 &  7.76$\pm$1.66 & 3.54$\pm$0.54 & $7.0\times 10^{16}$ & \multirow{3}{*}{$3.1\times10^{-5}$} \\
&  & \coc  & 70.46$\pm$0.05 &  1.29$\pm$0.06 & 2.14$\pm$0.11 & $3.2\times 10^{15}$ &  \\
&  & \hcop & 69.63$\pm$0.24 &  0.28$\pm$0.02 & 6.50$\pm$0.57 & $2.2\times 10^{12}$ &  \\ 
\hline
\multirow{3}{*}{3C\,396-M} & \multirow{3}{*}{10.3} 
   & \coa  & 69.67$\pm$0.09 &  6.94$\pm$0.29 & 3.82$\pm$0.27 & $6.4\times 10^{16}$ & \multirow{3}{*}{$5.6\times10^{-6}$} \\
&  & \cob  & 69.61$\pm$0.09 &  2.28$\pm$0.20 & 2.22$\pm$0.22 & $6.1\times 10^{15}$ &  \\
&  & \hcop & 69.81$\pm$0.12 &  0.16$\pm$0.02 & 1.86$\pm$0.29 & $3.6\times 10^{11}$ &  \\ 
\hline
\multirow{3}{*}{W51C-C2} & \multirow{3}{*}{13.4} 
   & \coa  & 61.72$\pm$0.06 & 10.00$\pm$0.15 & 8.26$\pm$0.14 & $2.4\times 10^{17}$ & \multirow{3}{*}{$2.3\times10^{-5}$} \\
&  & \coc  & 62.01$\pm$0.18 &  0.76$\pm$0.08 & 3.50$\pm$0.41 & $3.2\times 10^{15}$ &  \\
&  & \hcop & 62.49$\pm$0.22 &  0.38$\pm$0.02 &10.79$\pm$0.53 & $5.5\times 10^{12}$ &  \\ 
\hline
\multirow{3}{*}{HC\,40-W} & \multirow{3}{*}{10.2} 
   & \coa  & 36.01$\pm$0.03 &  6.90$\pm$0.07 & 5.32$\pm$0.06 & $7.8\times 10^{16}$ & \multirow{3}{*}{$3.6\times10^{-5}$} \\
&  & \coc  & 35.74$\pm$0.10 &  0.56$\pm$0.04 & 2.77$\pm$0.25 & $1.7\times 10^{15}$ &  \\
&  & \hcop & 35.36$\pm$0.04 &  0.69$\pm$0.02 & 3.27$\pm$0.09 & $2.8\times 10^{12}$ &  \\ 
\hline
\multirow{3}{*}{CTB109-E} & \multirow{3}{*}{13.8} 
   & \coa  & $-54.64\pm$0.01 &  10.39$\pm$0.10 & 1.23$\pm$0.01 & $2.5\times 10^{16}$ & \multirow{3}{*}{$1.0\times10^{-5}$} \\
&  & \cob  & $-54.69\pm$0.10 &   1.41$\pm$0.07 & 0.91$\pm$0.05 & $1.6\times 10^{15}$ &  \\
&  & \hcop & $-54.63\pm$0.07 &   0.14$\pm$0.01 & 1.36$\pm$0.16 & $2.6\times 10^{11}$ &  \\ 
\hline
\multirow{3}{*}{CTB109-N} & \multirow{3}{*}{15.0} 
   & \coa  & $-48.73\pm$0.01 &  11.56$\pm$0.06 & 2.83$\pm$0.02 & $9.6\times 10^{16}$ & \multirow{3}{*}{$1.7\times10^{-5}$} \\
&  & \cob  & $-48.55\pm$0.01 &   4.08$\pm$0.04 & 1.94$\pm$0.02 & $1.1\times 10^{16}$ &  \\
&  & \hcop & $-48.28\pm$0.02 &   0.54$\pm$0.01 & 2.01$\pm$0.06 & $1.6\times 10^{12}$ &  \\ 
\enddata
\tablecomments{
$a$: The excitation temperature $T_{\rm ex}$ are obtained from the optically thick \coa\ emission. It is assumed to be the same for different species. \\
$b$: Central velocities of the spectral components. \\
$c$: The abundance ratio between \hcop\ and \coa. The column densities of \coa\ are adopted directly from \coa\ instead of scaling from \cob\ and \coc. \\
$d$: The \hcop\ emission towards W30-S consists of two components. 
}
\end{deluxetable*}

\begin{figure*}[!t]
\centering
\includegraphics[width=0.98\textwidth]{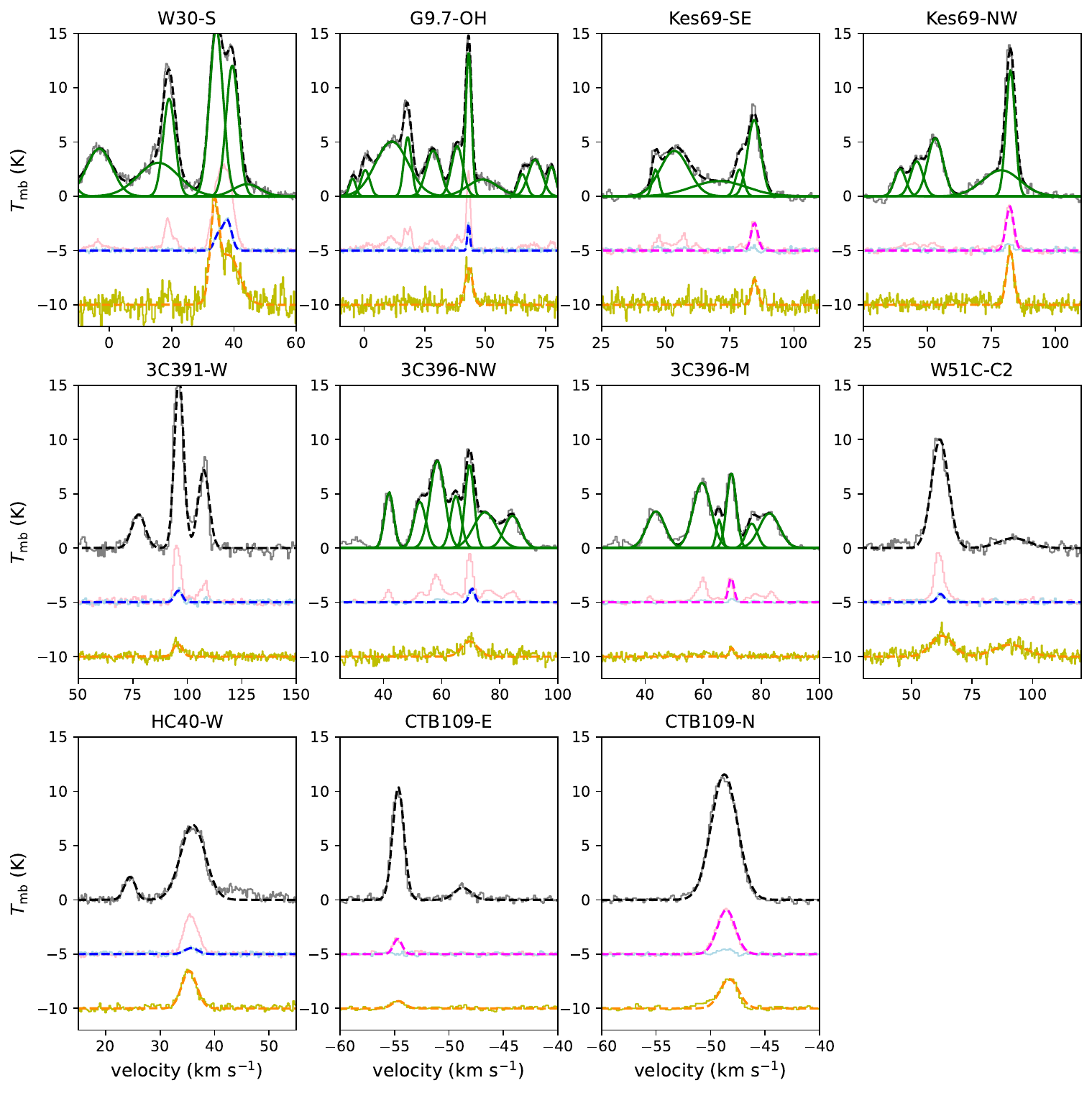}
\caption{Results of spectral decomposition of \coa, \cob, \coc, and \hcop\ towards 11 selected regions. 
The gray, pink, light blue, and yellow lines show the original spectra of \coa, \cob, \coc, and \hcop, respectively, while the dashed black, magenta, blue, and orange lines show the corresponding best-fit spectra. 
The spectra and \cob, \coc, and \hcop\ are lowered by 5\,K, 5\,K, and 10\,K for better visualization. 
The \hcop\ spectra are multiplied by a factor of 5. 
When the number of \coa\ components is greater than 3, the different velocity components are plotted in solid green lines. 
\label{fig:spec_fit}}
\end{figure*}

The \NhcopNco\ abundance ratio has been found to be enhanced in shocked regions of SNRs \citep{Zhou_Unusually_2022b,Tu_Shock_2024}. 
To study this abundance ratio towards the observed SNRs, we select 11 regions among the 20 regions whose spectra are shown in Figure \ref{fig:spec}. 
Since the \coa\ spectra may contain several velocity components, we discard 9 regions where the \coa\ spectra are too complicated to be decomposed and where the \hcop\ spectra has too low S/N to be fitted by Gaussian profiles. 
Towards 3C\,391-OH, the broadened \coa\ and \hcop\ may not trace the same shocked component \citep{Tu_Yebes_2024}, so we do not further investigate the emission from this region. 

\par

We fit the \coa\ spectra with multiple Gaussian velocity components. 
We refer to the peak positions of \cob\ and \coc\ to determine the number and positions of the \coa\ components. 
For the CO isotopes, we fit the \coc\ spectra when they exhibits high S/N and large linewidth ($\rm FWHM>5$ velocity channel width), otherwise we fit the \cob\ spectra. 
The procedure of the multi-Gaussian fitting is similar to what we did towards SNR W28 \citep{Tu_Shock_2024}. 
Hereafter we mainly focus on the components where \hcop\ emission is detected and consistent with the systemic velocity of the SNR. 

\par

The results of the spectral decomposition are shown in Figure \ref{fig:spec_fit} and Table \ref{tab:N}. 
The \hcop\ line towards W30-S consists of two velocity components, and both are shown. 
In several regions, broadened components of \coa\ are included (e.g. around +71 \kms\ towards Kes\,69-SE) to better fit the observed spectra, but theses may be due to the artifacts and do not necessarily mean shocked components. 
Some artifacts have also affected the fitting results of \hcop\ towards W30-S and 3C\,396-NW, resulting in broad line profiles. 
Towards G9.7-OH, the \hcop\ spectrum exhibits a double-peak profile (See Section \ref{sec:res_hcop}), but here we neglect it and fit the entire spectrum with one Gaussian component. 
Towards W51C-C2, the +61 \kms\ components of \coa\ and \hcop\ also exhibit broaden profiles, but they are caused by multiple velocity components that cannot be distinguished. 
We will further discuss the molecular clump W51C-C2 in a forthcoming paper with another observation (Tu et al., submitted). 
Towards CTB109-N, although we find possible line wing structures of \coa\ and \hcop\ (see Section \ref{sec:disc_broadening}), it is hard to distinguish so we regard the entire spectra as one single component. 

\par

To estimate the column densities of CO and \hcop, we follow the method in \citet{Mangum_How_2015} based on the local thermodynamic equilibrium (LTE) assumption. 
We also assume that the beam filling factor is 1. 
Since all the relevant \coa\ components are consistent with a \cob\ or \coc\ component, the \coa\ emission is expected to be optically thick, so the excitation temperature of the \coa\ lines can be estimated by: 
\begin{equation}
    T_{\rm ex} = \frac{h\nu/k}{ \ln{\left( 1+\frac{h\nu/k}{T_{\rm peak}+J_{\nu}(T_{\rm bg})} \right)} }, 
\end{equation}
where $J_\nu(T)=(h\nu/k)/[\exp{(h\nu/kT)}-1]$ is the Rayleigh-Jeans equivalent temperature and $T_{\rm bg}=2.73$ K. 
In the LTE condition, the excitation temperatures are assumed to be the same for different molecular transitions. 
The column densities of the species can then be estimated by: 
\begin{equation}
\begin{split}
    N = \left( \frac{3h}{8\pi^3S\mu^2R_i} \right) \left( \frac{Q_{\rm rot}}{g_Ig_Jg_K} \right) \left( \frac{\exp{(E_{\rm u}/kT_{\rm ex})}}{\exp{(h\nu/kT_{\rm ex})}-1} \right) \\
    \times \int -\ln\left[ 1 - \frac{T_{\rm mb}}{J_\nu(T_{\rm ex})-J_\nu(T_{\rm bg})}\right] \, dv, 
    \label{eq:N}
\end{split}
\end{equation}
where $Q_{\rm rot}\approx kT/hb+1/3$ is the rotational partition function and other parameters were explained in detail by \citet{Mangum_How_2015}. 
The estimated excitation temperatures and column densities are listed in Table \ref{tab:N}. 

\par

We note that the typical isotope ratios in the Galactic inner disk (2--6 kpc from the Galactic center) are $\rm ^{12}C/^{13}C\sim 40$ and $\rm ^{16}O/^{18}O\sim 327$ \citep{Yan_Direct_2023}. 
However, the isotope ratios of CO we obtained is significantly lower than these typical values (See Table \ref{tab:N}). 
This could be because the optical depths are underestimated in Equation \ref{eq:N}. 
For \coa, its 1--0 transition tends to be optically thick, so the column densities may be underestimated, while the estimated column densities of \cob\ and \coc\ are closer to the real values than \coa\ does because they are almost optically thin. 
The \hcop\ 1--0 line, although much weaker than the \coa\ 1--0 line, could also be optically thick even when the column density is not high ($\gtrsim 10^{12}\rm\ cm^{-2}$) \citep{vanderTak_computer_2007}. 
Therefore, the column densities of \hcop\ in Table \ref{tab:N} could also be underestimated. 
To reduce the uncertainty brought by the underestimation of optical depth in Equation \ref{eq:N} to some extent, we divide $N(\rm HCO^+)$ by the $N(\rm CO)$ directly obtained from the \coa\ spectra instead of using the CO isotopes scaled with the typical isotope ratios. 
However, this cannot fully eliminate the uncertainty on optical depth. 
Another source of uncertainty is the LTE assumption, which may be invalid at low volume densities. 
Therefore, our calculation of \NhcopNco\ can only be regarded as a rough order-of-magnitude estimation. 
The results of our estimation are listed in Table \ref{tab:N}. 
We find that the \NhcopNco\ in our selected regions are all of orders $\sim 10^{-5}$ or $\sim 10^{-6}$. 

\par

Enhanced \NhcopNco\ has been proposed to be a tracer of enhanced CR ionization rate according to chemical simulations \citep[e.g.,][]{Ceccarelli_Supernova-enhanced_2011,Bayet_Chemistry_2011}. 
However, \NhcopNco\ varies significantly from cloud to cloud, and we do not find difference between the MCs associated with the SNRs in our sample and typical quiescent MCs ($\sim10^{-4}$--$\sim10^{-6}$ \citep[e.g.,][]{Agundez_Chemistry_2013,Miettinen_MALT90_2014,Fuente_Gas_2019}), even if in the SNRs with evidence of enhanced CR ionization rates (e.g., W30, G9.7, Kes\,69, 3C\,391, 3C\,396, W51C, and CTB109 as listed in Table \ref{tab:info}). 
Similarly, in most SNRs with solid evidence of enhanced CR ionization rates, the \NhcopNco\ abundance ratios ($\sim 4\times 10^{-5}$ in W51C \citep{Ceccarelli_Supernova-enhanced_2011} and W28 \citep{Vaupre_Cosmic_2014}, and $\sim 8\times 10^{-5}$ in 3C\,391 \citep{Tu_Yebes_2024}) do not deviate from the typical values in quiescent MCs. 
In W51C and W28 where the enhanced CR ionization rates were estimated with both \NhcopNco\ and $N({\rm DCO^+})/N({\rm HCO^+})$ \citep{Ceccarelli_Supernova-enhanced_2011,Vaupre_Cosmic_2014}, the results were mainly deduced from the unusually low column densities of $\rm DCO^+$ instead of enhanced \NhcopNco. 
Therefore, it appears that \NhcopNco\ is not an effective tracer of chemistry induced by CRs. 
Significantly enhanced \NhcopNco ($\sim 10^{-3}$) has only been found in SNR W49B \citep{Zhou_Unusually_2022b}, but this high ratio was discovered in the shocked MC with broadened molecular line profile where shock chemistry cannot be ignored. 
Another example is SNR W28, where the \NhcopNco\ is enhanced in shocked MC ($\sim 10^{-4}$) compared with unshocked MCs close to the SNR ($\sim 10^{-5}$) \citep{Tu_Shock_2024}. 
It is likely that shock interaction is important for CRs to enhance the \NhcopNco\ abundance ratio.

\par

We note, from Table \ref{tab:info}, that in most SNRs with signature of enhanced CR ionization rates, \hcop\ emission is detected (W30, G9.7, Kes\,69, 3C\,391, Kes\,78, W51C, and CTB109). 
This suggests that the detection of \hcop\ indicates the presence of the target MCs to produce hadronic $\gamma$-rays. However, there are cases where \hcop\ is detected with no evidence of enhanced CR ionization (HC40) and where no \hcop\ is detected but solid evidence of enhanced CR ionization has been observed (G16.7+0.1). 
Therefore, whether \hcop\ can be detected depends strongly on the local environment of the SNR, especially the gas density and temperature, instead of whether the CR ionization rate is enhanced to elevate the abundance of \hcop.

\par

X-rays from the SNRs can also lead to enhanced ionization rates in MCs \citep{Maloney_X-Ray--irradiated_1996,Viti_Molecular_2017}. 
However, according to the estimation of \citet{Tu_Yebes_2024}, the X-ray ionization rate induced by SNR 3C\,391 is negligible. 
Since 3C\,391 is a bright SNR in the X-ray band and the other SNRs we observe do not show significantly high X-ray luminosities \citep[e.g.,][]{Albert_exploration_2022a}, we consider that X-ray ionization in the target SNRs can be ignored.

\section{Conclusion} \label{sec:con}
In this paper, We present our observation of \hcop\ and \hcn\ 1--0 lines towards 13 SNRs interacting with MCs with the PMO 13.7m telescope, supplemented by archival data of CO isotopes. 
The results can be summarized as follows: 

\par

1. We find strong emission of \hcop\ towards the FOVs of SNRs W30, G9.7$-$0.0, Kes\,69, 3C\,391, 3C\,396, W51C, HC\,40 and CTB109 in the velocity intervals proposed to show evidence of SNR-MC interaction. 
Weak \hcop\ emission is detected in Kes\,78, while no \hcop\ emission is found in G16.7$+$0.1, 3C\,397, Kes\,75 and CTB87. 

\par

2. We find a \coa\ arc surrounding the radio continuum of SNR G9.7$-$0.0 in the northern, eastern and southern part, with an expansion as revealed by the position-velocity diagram around the LSR velocity of the 1720 MHz OH maser. 
This bubble is likely to be driven by the stellar wind of the SNR progenitor. 
With the new radio data, we also find a \coa\ arc spatially coincident with the northwestern radio arc of Kes\,69. 
This suggests that, in addition to the known molecular shell along the southeastern boundary
of the SNR, there is also a molecular arc along the northwestern boundary.

\par

3. Significant shock broadening is found in SNRs 3C\,391 and W51C, and CTB109 exhibits a possible blue-shifted line wing brought by shock interaction. 
The line profile towards CTB109-N is consistent with the X-ray absorption. 
This non-detection of line broadening in other SNRs may be due to the limited sensitivity and angular resolution of our observation, as well as the possibility that the timescale of the interaction is too short to allow for detectable line broadening. 
For the 1720 MHz OH masers towards G16.7$+$0.1, the northeast of Kes\,69, the northeast of 3C\,391, and Kes\,78, we did not find corresponding \hcop\ or \hcn\ emission, probably because the OH masers only take up a rather small physical scale and cannot represent the physical properties of the entire MC. 

\par

4. The median values of \IhcopIhcn\ in all of the SNRs except Kes\,69 fall in 0.65--1.0. 
The highest values and largest scatter of \IhcopIhcn\ are found in W51C, which may be caused by self-absorption which is affected by expansion motion and is more severe for \hcn. 
We do not find significant variation of \IhcopIhcn\ between broad-line and narrow-line regions, and among different SNRs. 
The obtained \IhcopIhcn\ also deviate little from typical values found in Galactic MCs. 
The observed \IhcopIhcn\ line ratio results from a complex interplay of excitation and chemical effects, so we caution on using \IhcopIhcn\ line ratio as a diagnostic of SNR feedback and CR ionization. 

\par

5. We estimate the \NhcopNco\ abundance ratio in 11 regions towards the observed SNRs. 
The abundance ratios \NhcopNco\ in all of the selected regions are of orders $\sim 10^{-5}$ or $\sim 10^{-6}$, which is similar to the values in typical quiescent MCs. 
Combining the \NhcopNco\ ratio towards the points with enhanced CR ionization rates in W28, W51C and 3C\,391, we suggest that the \NhcopNco\ ratio may not be an effective tracer of CR-induced chemistry.

\begin{acknowledgments}
The authors thank the anonymous reviewer for helpful suggestions. 
The authors are grateful to the staff of Qinghai Radio Observing Station at Delingha for their help during the observation. 
T.-Y. Tu thanks Ping Zhou for helpful discussion. 
Y.C. acknowledges the support from NSFC grants Nos. 12173018, 12121003, and 12393852.

\end{acknowledgments}

\vspace{5mm}

\facilities{PMO:DLH, No: 45m, VLA}

\software{astropy \citep{AstropyCollaboration_Astropy_2018, AstropyCollaboration_Astropy_2022},  
          Spectral-cube \citep{Ginsburg_Radio_2015}, 
          GILDAS (Gildas Team, \url{https://www.iram.fr/IRAMFR/GILDAS/}), 
          Montage (\url{http://montage.ipac.caltech.edu/}, 
          Matplotlib (\url{https://matplotlib.org}))
          }





\bibliography{SNR_HCO+_article}{}
\bibliographystyle{aasjournal}

\end{CJK*}
\end{document}